\documentclass[12pt, a4paper, nohyper, notoc]{JHEP3}
\usepackage[intlimits,tbtags]{amsmath}
\usepackage{amsfonts}
\usepackage{amssymb}
\newcommand{\bes}{\begin{eqnarray*}}
\newcommand{\ees}{\end{eqnarray*}}

\newcommand{\bet}{\beta_{H}}
\newcommand{\be}{\begin{equation}}
\newcommand{\ee}{\end{equation}}
\newcommand{\ber}{\begin{eqnarray}}
\newcommand{\eer}{\end{eqnarray}}

\newcommand{\lp}{\left(}
\newcommand{\rp}{\right)}
\newcommand{\lk}{\left\{}
\newcommand{\rk}{\right\}}
\newcommand{\lc}{\left[}
\newcommand{\rc}{\right]}
\newcommand{\dif}{\mathrm{d}}
\newcommand{\2}{\,2}
\newcommand{\bh}{\beta_{\scriptscriptstyle{H}}}   
\title{On the stability of the primordial
closed string gas}

\author{Manuel A. Cobas, M. A. R. Osorio, Mar{\'{\i}}a Su\'arez\\
Dpto. de F{\'{\i}}sica, Universidad de Oviedo\\
Avda. Calvo Sotelo 18\\
E-33007 Oviedo, Asturias, Spain\\
\email{cobas, osorio, maria@string1.ciencias.uniovi.es}}

\preprint{\hepth{0507088}}

\abstract{We recast the study of a closed string gas in a toroidal 
container in the physical situation in which 
the single string density of states is independent 
of the volume because energy density is very high. 
This includes the gas for the well known Brandenberger-Vafa cosmological scenario. 
We describe the gas in the grand canonical and microcanonical ensembles. 
In the microcanonical description, we find a result that clearly 
confronts the Brandenberger-Vafa calculation
to get the specific heat of the system. The important point is that we use
the same approach to the problem but a different regularization. By the way, we
show that, in the complex temperature formalism, at the Hagedorn singularity,
the analytic structure obtained from the so-called F-representation of the free
energy coincides with the one computed using the S-representation.}

\begin{document}

\section{Introduction}
Treating a gas of free closed superstrings when all the spatial dimensions are
closed and the system is kept in a thermal bath
leads us to the well known conclusion that the Helmholtz free energy diverges as
one approaches the Hagedorn
temperature from the low temperature regime. This way, as the energy
$U$ also diverges, one can conclude that, in a fixed temperature description,
the Hagedorn temperature is a maximum one for the system. Let us recast in
the next section what the details
for the description in the macrocanonical ensemble with $\mu=0$ are. In
particular,
we will remind to the reader that energy fluctuations give us a
measure about the very possibility
of a fixed temperature description of the system. In section 3 we will compare
our results with the thermodynamics that stems from the generalized ensemble
description of an analogous
extensive system. After all this, we will have then got some results to get into
the treatment of the system at fixed energy in section 4. There, in a first
subsection, we will
remember and critically recast the well known computation by Brandenberger and
Vafa \cite{vafa}. Next, 
we will present another computation that  will reinforce the conclusion
that the specific heat, as a function of energy, is divergent. A final
subsection will be devoted to explain whether  volume dependent 
corrections can change the picture. Finally, section 5 
will present a few comments and serve as a reminder of the main results. 

\section{The macrocanonical description of closed strings at finite size} 

In a macrocanonical description at null chemical potential, 
the grand canonical partition function can be equated with the canonical
partition function at a given number of strings $N^*$. This number maximizes
the canonical partition function $Z(\beta,R,N)$ (i.e.,
$\left[\partial\,Z/\partial\,N\right]_{N=N^*}=0$). When
the fluctuations in the number of strings are small, this
maximum coincides with the averaged number of strings, $\overline{N}$, as
computed in the grand canonical ensemble. To be more concrete: $\Theta\lp
\beta, R, \mu=0\rp = \sum_{N=0}^{\infty}\,Z\lp \beta, R, N\rp \approx Z\lp
\beta, R, N^{*}\lp \beta, R\rp\rp$ and $Z$  finally  results a function of
$\beta$ and $R$ only. This $Z$ is what we call the partition function
for the system of free strings. Minus the logarithm of the partition function
divided by $\beta$ is what we call the free energy of the system and it is
only a function of $R$ and $\beta$ and not of the number of strings (see
\cite{exten}). This will exactly correspond to the thermodynamical free
energy whenever a thermodynamical limit can be defined. The computation 
for the gas of free superstrings gives

\begin{multline}
\label{closedF}
-\beta F\lp\beta\rp = \beta\,\frac{2^{\,6}}{\pi\sqrt{\alpha'}}\,
\int_{0}^{+\infty}\,\dif\tau_2\,\tau_2^{-3/2} \,\theta_2\lp 0, 
\frac{\mathrm{i}\beta^{\,2}}{\pi^{\,2}\tau_2\alpha'}\rp
\int_{-1/2}^{1/2}\,\dif\tau_1\,\left|\theta_4\lp 0,
2\tau\rp\right|^{\,-16}\\
\times\sum_{\vec{m},\vec{n}}\,
\mathrm{e}^{-\pi\tau_2\lp\frac{R^{\,2}}{\alpha'}\vec{m}^{\,2}  +  
\frac{\alpha'\vec{n}^{\,2}}{R^{\,2}}\rp +
2\pi\mathrm{i}\tau_1\vec{m}\cdot\vec{n}}
\end{multline}

This is the free energy in the  so called $S$-representation. It is
the result  one gets when computing the Helmholtz free energy in the light-cone
gauge or  by summing up over the field content of the string
(analog model)
(see  \cite{alvarezmar}).
In the conclusions we will comment more on this point and  its relation
to the $F$-representation, that coincides with the computation of the free
energy as a vacuum energy for the Euclidean theory  with Euclidean time
of length $\beta$ including winding modes around it.

Now, let us introduce an ultraviolet dimensionless
cutoff $\epsilon$ in $\tau_{\,2}$. This automatically produces the splitting of
the free energy as $F\lp \beta\rp = F^{\,l}\lp\beta\rp + F^{\,h}\lp\beta\rp$
where $F^{\,l}$ is got\footnote{$F^{\,l}$, as it is written in this proper time
representation,
shows a divergence when $\tau_{\,2} \rightarrow +\infty$ for the massless 
excitations of the string when momentum and winding numbers vanish. This
is an artifact of this representation that results from the second quantization
of the vacuum state then producing a divergence as $\mathrm{ln} \tau_{\,2}$ when
$\tau_{\,2}$ goes to infinity. This divergence should not be present with
finite volume or, more precisely, when the momenta are not dense, and can be
subtracted without affecting the results of our work.} by integrating
$\tau_{\,2}$ from $\epsilon$ to $+\infty$
and $F^{\,h}$  by integrating the same integrand over $\tau_2$ from $0$
to $\epsilon$.

The integral over $\tau_1$ simply represents the left-right
level matching condition, i.e., a Kronecker delta of the form
$\delta_{N -\tilde{N}+\vec{m}\cdot\vec{n}, \,0}$. Here, $\tilde{N}$ and $N$
stands for the right and left oscillator numbers. 

Supposing $\epsilon \ll 1$ has various implications. The Jacobi
$\theta_2$ function, having the $\beta$ dependence, can then be approximated by
the leading terms (two in fact) in the series expansion representing it. This is physically
equivalent to taking the classical  (Maxwell-Boltzmann) statistics approximation. 
 One can also use the fact that $\tau_2 \leq \epsilon$ to approximate
with arbitrary precision the contribution from integrating over $\tau_1$,
i.e., implementing the left-right level matching condition.

To compute first the integral over $\tau_1$, it is very useful to note that,
when $\tau_2$
is small, the main contribution to the integrand will come
from a neighborhood of $\tau_1=0$. 

Furthermore,
one can choose $\epsilon$ small enough so as to have that $\epsilon
\alpha'/R^{\,2} \ll 1$ and, simultaneously, $\epsilon R^{\,2}/\alpha' \ll 1$.
It is enough to choose $\epsilon$ to fulfill the most stringent criterion. In fact, one 
criterion converts into
the other by T-duality ($R\rightarrow \alpha'/R$). If $R \gtrsim \sqrt{\alpha'}$,  
one can choose $\epsilon \ll \alpha'/R^{\2}$ in order to accomplish both criteria. 
When $R \lesssim \sqrt{\alpha'}$  we have that $\epsilon \ll R^{\2}/\alpha'$ 
suffices to satisfy both relations. We will  see how these criteria,
 that define the physical system
we are treating and the kind of thermodynamical limit we are taking, 
can be expressed in terms
of energy density. This
allows us to approximate the sum
over $\vec{m}$ and $\vec{n}$ (i.e., over winding and momentum numbers) by a
multiple integral over the whole $\mathbb{R}^{\,18}$ on the real variables
$m_1,...,m_9,n_1,...,n_9$. This is a valid approximation as given by the
Euler-Maclaurin formula (see \cite{nuestro}). The  multiple integral over
windings and momenta shows us that, for the contribution coming from the
ultraviolet degrees of freedom to $F\lp\beta\rp$ (i.e., the contribution
encoded in $F^{\,h}$), no dependence on $R$, and then on the volume $V=(2\pi
R)^{\,9}$, will survive. This is so because the change of variables  of unit
Jacobian $\vec{m}\longrightarrow \vec{m}\sqrt{\alpha'}/R$, 
$\vec{n}\longrightarrow \vec{n} R/\sqrt{\alpha'}$ makes $R$ to disappear in
the multiple integration.  Everything happens as putting
$R=\sqrt{\alpha'}$ that simply shows the fact that  a situation
in which $R$ is in a neighborhood of $\sqrt{\alpha'}$ also suffices to compute
the sum by
an integral. This will be linked to the cosmological
situation in which all the spatial dimensions are of the order of the selfdual
length and the system is, in some sense, small.

With all this together one can finally write

\be
\begin{split}
I(\tau_2)\equiv 2^{\,8}\,&
\int_{-1/2}^{+1/2}\,\dif\,\tau_1\,\left|\tau\right|^{\,8}
\,\left|\theta_2 (0,-1/(2\tau))\right|^{\,-16}
 \int_{\mathbb{R}^9\times\mathbb{R}^9}\, \dif\vec{l}\,\dif\vec{k}\, 
 \mathrm{e}^{-\pi\tau_2(\vec{l}^{2} + \vec{k}^{2})}\mathrm{e}^
{2\pi \vec{l}\cdot\vec{k}\tau_1}
\\ &=
2^{\,-17/2}\,\tau_2^{\,1/2}\,\mathrm{e}^{\,2\pi/\tau_2}\,\sum_{i=0}^{+\infty}\,a
_ i \,\tau_2^{\,i}
\end{split}
\label{tau1}
\ee
Where the modular properties of the transverse partition function have been
used. We remark again, that the main contribution in the $\tau_2
\rightarrow 0$ limit  for the $\tau_1$ integral comes from the neighborhood of
$\tau_1=0$. The coefficients $a_i$ are computable numbers (see
the Appendix). In particular, $a_0=1$

$F^{\,h}\lp \beta\rp$ can then be approximated, for 
$\2\pi\,\lp\beta^{\,2}-\beta_H^{\,2}\rp\ll 
\epsilon\beta_H^{\,2}$ ,
as\footnote{The well known relation $\Gamma \lc a+1, x \rc = a \Gamma\lc a,x\rc
+ x^{\,a}\mathrm{e}^{\,-x}$ is very useful to this purpose.}

\be
\label{libre}
-F^{\,h}\lp\beta\rp \approx \,\Gamma\lp 0,\frac{2\pi\,\lp\beta^{\,2}
-\beta_H^{\,2}\rp}{\epsilon\beta_H^{\,2}}\rp\,
\sum_{n=0}^{+\infty}\,b_n
\lp\frac{2\pi\,\lp\beta^{\,2}-\beta_H^{\,2}\rp}{\beta_H^{\,2}}\rp^n
\ee
$\beta_H = \pi \sqrt{8\,\alpha'}$ is the inverse Hagedorn temperature for closed
superstrings type IIA and IIB (both have the same free energy because are
indistinguishable at finite $T$) and
$\beta-\beta_H > 0$.
The $b_n$ are coefficients that can be directly connected to the $a_i$; for
example, $b_0 = 1/\beta_H$, and, in general, the $b_n \propto 1/\beta_H$
are independent of $\epsilon$ computable numbers. The important
point now is that, as discussed in the Appendix, 
 
\be
\sum_{n=0}^{+\infty}\,b_n
\lp\frac{2\pi\,\lp\beta^{\,2}-\beta_H^{\,2}\rp}{\beta_H^{\,2}}\rp^n
=\sum_{n=0}^{+\infty}\,(-1)^n \frac{\lp\beta -\beta_H\rp^n}{\beta_H^{\,n+1}} =
\frac{1}{\beta}
\label{magic}
\ee

 As it is implied by the behavior of $\Gamma[0, z]$ for big $z$,
$F^{\,h}$ goes
exponentially to zero as $\beta-\beta_H$ grows. Because there is no dependence
on the volume, it is not true that we can write $F\lp\beta\rp = -PV$. This
gives more importance and justifies the detailed way we have introduced
$F\lp\beta\rp$ at the beginning of this section. It is now clear from
the behavior of $\Gamma[0,z]$ when $z$ goes to zero that the contribution of
$F^{\,h}$ to the free energy when  $\beta\sim\beta_h$ is  much bigger than that of
$F^{\,l}$ because $|F^{\,h}|$ grows unbounded 
as $\beta \rightarrow \beta_H^+$ as long as
$F^{\,l}$  gets the finite value $F^{\,l}(\beta_H)$.  

The concrete behavior
around $\beta_H$ can be made more explicit 
by using that $\beta^{\,2} -\beta_H^{\,2}=
2\beta_H(\beta - \beta_H) +  (\beta - \beta_H)^{\,2}\approx 
2\beta_H(\beta - \beta_H)$.
One finally gets

\be
\begin{split}
-\beta  F^{\,h}\lp\beta\rp &\approx 
\int_{0}^{+\infty}\,\dif E \,\,
\theta\lp E - \frac{4\pi}{\epsilon\beta_H}\rp\, \mathrm{e}^{-\,\beta E}
\,\frac{\mathrm{e}^{\,\beta_H\,E}}{E}\\ &=
\Gamma\lp 0, \frac{4\pi\lp\beta
-\beta_H\rp}{\epsilon\beta_H}\rp
\label{sinnombre}
\end{split}
\ee
This, by using inverse Laplace transformation,
 easily provides us the main ingredient to get the fixed energy
description,  $\Omega_1(E)$, i.e., the single string density of states.

\be
\Omega_1^{\,h}\lp E\rp = \theta\lp E-\Lambda\rp\, 
\frac{\mathrm{e}^{\,\beta_H\,E}}{E} 
\label{single}
\ee

The step function shows the utility of the dimensionless $\epsilon$ parameter by
imposing the condition $E > \Lambda=4\pi/(\epsilon\beta_H)$. This, when
$R\gtrsim\sqrt{\alpha'}$,
finally enforces $E\gg 4\pi R^{\2}/(\beta_H\alpha')$ as the condition for the
validity
of \eqref{single}\footnote{It is important to remark
that $E$ is here the energy of the states which are accessible by one string. 
This condition can be compared
to the one in \cite{nuestro} for the validity of converting 
the sum over winding and momenta into an integral
in the direct calculation of $\Gamma_1$ from its very definition
($\Omega_1 = \dif \Gamma_1/\dif E$);
 the condition  is $E \gg R/\alpha'$. It is clear that
$E\gg 4\pi R^{\2}/(\beta_H\alpha')$ implies
 $E \gg R/\alpha'$ if $R\gtrsim\sqrt{\alpha'}$.}.
 For the T-dual situation, one gets the dual condition.

It is now an immediate task to get $U\lp\beta\rp$ around $\beta_H$

\be
U^{\,h}\lp\beta\rp = \frac{\partial\lc\beta F^{\,h}\lp\beta\rp\rc}
{\partial\beta} \\= \frac{1}{\beta -\beta_H}
\mathrm{e}^{-4\pi\lp\beta-\beta_H\rp/\lp\epsilon\beta_H\rp}\\
 \approx  \frac{1}{\beta-\beta_H}
\label{energy}
\ee
Where we have taken that $0 < \beta-\beta_H \ll
\epsilon\beta_H/(4\pi)=1/\Lambda$.

Fluctuations in the macrocanonical energy are very useful for our study.
They can be computed to give, near $\beta_H$

\be
\label{fluc}
\frac{\lc T^{\2}\,C_V^{\,h}(T, V)\rc^{1/2}}{U^{\,h}} =
1 + \mathrm{O}\lc\lp\beta-\beta_H\rp^{\,1}\rc
\ee

So energy relative  fluctuations are finite and not negligible. This means that
we may
expect the fixed energy ("micro") description and this macrocanonical
picture  not to be equivalent.

 The  entropy can easily be computed  as a function of $\beta$ to give

\be
S^{\,h}\lp\beta\rp=\beta^{\2}\,\frac{\partial F^{\,h}}{\partial\beta}
\approx \frac{\beta_H}{\beta-\beta_H} - \mathrm{ln}\lc
\Lambda\lp\beta-\beta_H\rp\rc 
\label{entropy}
\ee

The fundamental relation in the entropic representation can
now easily be obtained to be

\be
S^{\,h}  \approx \beta_H \,U^{\,h} + \mathrm{ln} U^{\,h}
\label{entropic}
\ee

In the entropic representation it is manifest 
that the non extensive
term $\mathrm{ln}\,U^{\,h}$ is the one responsible for the positivity of the
specific heat. When energy is high, the logarithm 
of the averaged energy is very
small as compared to energy itself. If one sees the big fluctuations as giving
the error of the energy variable to produce $S(\overline{E}=U)$, one perhaps
should neglect the logarithmic term because is smaller than the error. 

The macrocanonical calculation  gives us the number of strings as
 $-\beta F\lp \beta\rp$ if Maxwell-Boltzmann statistics
applies. Namely

\be
-\beta F^{\,h}\lp\beta\rp=
\overline{N}^{\,\,h}  \approx  -\mathrm{ln}\lc\lp\beta-\beta_H\rp\Lambda\rc 
 \approx\mathrm{ln}\,U^{\,h}
\label{numero}
\ee

One can write the entropy in terms of the number of strings as a function of
energy to get

\be
 S^{\,h} \approx \overline{N}(U^{\,h}) + \mathrm{e}^{\overline{N}(
U^{\,h})}
\ee

This expression for  the entropy can be compared to the one for a
regular system for which the entropy scales with the number of objects 
as in the black body, for instance.
In our gas we have that the entropy near $\beta_H$ (that also gives high $U$)
grows  exponentially with the 
number of objects.
This behavior in terms of the number of strings 
can also be compared to the open superstring case in which $T_H$ is a 
true maximum temperature 
\cite{openss}. For it, the entropy
grows as $2\overline{N}(U^{\,h}) +  K \overline{N}^{\,2}(U^{\,h})/V$, 
with $K$ a constant,
and is a degree one homogeneous function 
of energy and volume (extensivity is a property of the gas of open superstrings in the infinite
volume limit). 

It is then clear that the specific heat, as a function of the temperature,
 is positive for the high temperature phase, i.e., when
$T$ is near $T_H$. The problem is that the order one energy fluctuations tell
us that $U^{\,h}$ is
a bad canonical average for the high energy of the non-isolated  system.

\section{The generalized ensemble and extensivity}

The description at fixed temperature  we have made is one very special. 
The reason is that our system does not depend 
on volume because this variable
does not appear in the description of the system which is also at
 fixed temperature and null chemical potential.
 
A description through a generalized ensemble 
is one in which the system is characterized
 by intensive parameters. In a simple system they are 
pressure, temperature and chemical potential instead of
volume, energy and the number of objects which are the corresponding extensive parameters.
Since our string gas is one for which no volume dependence 
appears\footnote{This  is different from being a problem at zero pressure. In a system at
zero pressure, the volume would be a function of temperature. In our case, 
there is no volume dependence at all and then the pressure vanishes.}, the
ensemble we have named grand canonical is really a generalized ensemble.

This ensemble does not appear thoroughly 
treated in regular textbooks on Statistical 
Mechanics but can be found in \cite{hill2}. The main point to take into account
when reading 
this textbook 
is that the treatment of the description using this ensemble depends on the fact
that extensivity 
is  assumed for the system.
The first notable fact about this ensemble, if extensivity is assumed, 
is that fluctuations in 
volume, energy and the number of objects
must be big. The reason is that, in this picture, the system is characterized
 by pressure, temperature and $\mu$ and 
then volume, total energy and the number of objects can get any value with equal probability. 
It is worth to remark that this must be so when extensivity holds. 

Assuming extensivity in our problem
 would imply $\beta F\lp\beta\rp = 0$ because Euler's relation ($U-TS+PV-\mu N=0$)
 would hold and we have $\mu=0$ and no volume dependence. On the contrary,
non extensivity allows a non vanishing free energy as computed previously, 
the relative energy  fluctuations to be big (order one)
and, simultaneously, the fluctuations in the  number of strings 
to be small when energy is big enough.
Indeed, the fluctuations in $\overline{N}(T)$
 are small because, when Maxwell-Boltzmann statistics and the dilute gas
approximation hold, they are given for 
any system by
 
\be
\frac{\sqrt{\overline{\Delta N^{\2}}}}{\overline{N}(\beta)}= 
\frac{\sqrt{\lp z\partial_{z}\rp^{\2}\lp z\,q\rp}}
{q}= \frac {1}{\sqrt{q}}=\frac{1}{\sqrt{\overline{N}\lp\beta\rp}} 
\ee
 where $q=-\beta F\lp\beta\rp$ is the single object partition function, a function of 
 $T$ and $V$ in general and only of $T$ in our problem. For
the classical counting, $q$ gives  the number of objects. 
The general application of this result is not in contradiction with the fact
that, in 
the generalized ensemble when extensivity holds, the fluctuations in  the number
of objects
are big.  When  the system is extensive all the subsystems in the generalized ensemble
with different object numbers 
are equally probable and then, to get the partition function,  the sum over
the number of objects does not run up to infinity
(see \cite{hill2} again).

On the other hand, the fact that for any extensive system at $\mu=0$
and with no volume dependence one has $U-TS=0$ has immediate consequences.
By putting $T=\partial U/\partial S$, $U-S\, \partial U/\partial S = 0$ can be
understood as a differential equation 
 that can be solved to give $S = \beta_0 U$ where $\beta_0$
is a constant of integration that gives the constant inverse temperature of the
system. In our problem 
this $\beta_0$ is $\bh$. From
the point of view of  Legendre transformations, this is the most simple
and extreme case for which the transformation cannot be defined because the system has
an infinite specific heat. Furthermore, if we now associate a 
density of states to the obtained entropy
we have  $\Omega\lp E\rp = K\beta_0 \mathrm{e}^{\beta_0 E}$, where $K$ is a
positive dimensionless constant. 
Computing now back the partition
function $Z\lp\beta\rp$ thorough Laplace transformation we obtain\footnote{
We could have  included
a cutoff $\phi$ to get $Z\lp\beta\rp = K\,
\beta_0\mathrm{e}^{-\phi\lp\beta -\beta_H\rp}/\lp\beta -\beta_0\rp$.}
$Z\lp\beta\rp =K\,\beta_0 \int_{0}^{+\infty}\,\dif E\,\,\mathrm{e}^{-\lp\beta -
\beta_0\rp\,E}  =K\,
\beta_0/\lp\beta -\beta_0\rp$. So we find the surprising result that the free
energy is not
zero, but actually $\beta F\lp\beta\rp= \mathrm{ln}\lc\lp \beta -
\beta_0\rp/\lp K\beta_0\rp\rc$.
 The free energy must be negative and then 
has  physical meaning when $\beta$ is near and bigger than $\beta_0$ so,
after all, there is a maximum temperature $T_0$ and there must 
also be
an energy cutoff for the validity of $\Omega\lp E\rp$.
Finally, one exactly gets the singular  dominant term in 
the incomplete Gamma function in \eqref{sinnombre} identifying $T_0=T_H$; i.e.
the logarithmic contribution.
This is a very simple example of  non equivalence of 
ensembles so brightly explained in \cite{touchet} and that
relies upon the interplay between Legendre and Laplace transformations. It is
very important
to notice that the single object partition function $-\beta F\lp\beta\rp$ we get
actually  coincides with the one in \eqref{numero}. 
 
Conversely, one can easily 
show that if the  canonical energy is a function of $T$ such that, at $T=T_0$, 
$\lp\beta-\beta_0\rp\,U\lp\beta\rp$ gives zero (or a constant) when $\beta$
approaches
$\beta_0$ then, 
since $S(\beta)= \beta U(\beta) $ for a volume independent 
extensive system at 
$\mu=0$, one gets that $S(\beta) = \beta_0 U(\beta)$ around $\beta_0$. But, in
fact,
we have already arrived at the fundamental thermodynamic relation
$S=\beta_0\,U$ that implies
that the temperature is constant and equal to $T_0$. This contradicts the
hypothesis assuming that the internal energy is a function of a variable
$\beta$.
The final output is then that $S=\beta_0 U$ only makes sense in microcanonical
thermodynamics and, more than this, extensivity holds for the microcanonical
thermodynamics we get.

It is certainly a notorious fact 
that, in this particular case, one can deduce the exponential growth of the 
density of states with energy from the hypotheses
that extensivity holds, the system does not depend on volume  and  equilibrium
is got at zero chemical potential. In fact, all the above reasoning simply tells
us 
that $ \Omega(E)$, the density of states for the gas of strings, is
a constant times 
$\mathrm{e}^{\beta_H\,E}$ when energy goes beyond a certain value, let's call it
$\phi$. However, this
contradicts the computation in \cite{vafa} in which a different non independent
of the volume non extensive high energy entropy is found. This makes us suspect
that there is something wrong in that computation.

On the other hand, the grand canonical description of 
the closed string gas
near $T_H$ and under the condition that windings and momenta are equivalent
is 
really a generalized ensemble description but without assuming extensivity. 
Non extensivity is usually
related to a kind of smallness of the system in size or number of objects, 
the presence of long range interactions or a critical behavior. We will dwell 
a bit more on this point in the final section.

\section{The fixed energy description of closed strings at finite size}

In order to study the fixed energy description of the gas of closed strings,
an expression for the multiple string density of states, $\Omega\lp E\rp$, is needed. This
has been done in the past in several ways. One could use, for example, a
saddle point approximation to calculate the asymptotic expression for the
density of states of the string gas, using the fact that $\Omega\lp E
\rp={\cal{L}}^{-1}\lk\exp\lp -\,\beta F\lp \beta\rp\rp\rk$. However, once the single
string density of states is provided, $\Omega\lp E\rp$ can be obtained
using the convolution theorem \cite{convo}. This is the method used
long time ago by R. Brandenberger and C. Vafa 
\cite{vafa} (see also \cite{frautschi}), 
who calculated\footnote{Only for the Maxwell-Boltzmann statistics $\Omega_{n}$ 
can be understood as the density of states for a gas with $n$ strings \cite{exten}.}
 
\ber
\Omega_{n}\lp E\rp=\frac{1}{n!}\int_{\Lambda}^{E}\prod_{i=1}^{n}\dif
E_{i}\Omega_{1}
( E_{i})\delta\lp\sum_{i}E_{i}-E\rp\label{omegabase}.
\eer 
$\Omega_1(E)$ can be expressed as sum of two terms $\Omega_1^l(E)$
and $\Omega_1^h(E)$ where the superscripts $l$ and $h$ refer to the low and high
energy expressions.  This introduces a cutoff  $\Lambda$  separating both
regimes that can  exactly be written as our  cutoff $4\pi/(\epsilon\beta_H)$ 
on the energy of one string. 
 In fact we can obtain $\Omega_N(E)$ with a high degree of
accuracy only from the convolutions of
$\Omega_1^h(E)$ as in the gas  of open strings \cite{openss}.  The
convolution between $\Omega_1^l(E)$ and $\Omega_1^h(E)$ to give a
contribution to $\Omega_2^h(E)$ is negligible, as can clearly be seen in Fig.~\ref{F1} 
where the comparison between the exact calculation\footnote{
This means that we have used an exact form for the coefficients giving the
degeneration number at each mass level of the superstring 
and the sum over winding and momenta
has been obtained by integration.} of
$\omega_2(E,t)=\Omega_1(E-t)\,\Omega_1(t)$ and its  approximation resulting from
considering only the high energy part approximated by $\Omega_1^{\,h}$ in \eqref{single}
is shown at $E=56$, $R=\sqrt{\alpha^{'}}=1$.  $\omega_2$ is a measure of
equipartition of energy  between
the two strings. In fact, it is Fig. \ref{F1} that lets us fix
the cutoff 
$\Lambda=\frac{4\pi}{\epsilon\beta_H}$

\FIGURE{
\let\picnaturalsize=N
\def\picsize{3.0in}
\def\picfilename{convoaltaalta1.eps}
\ifx\nopictures Y\else{\ifx\epsfloaded Y\else\input epsf \fi
\let\epsfloaded=Y
\centerline{\ifx\picnaturalsize N
\epsfxsize \picsize\fi\epsfbox{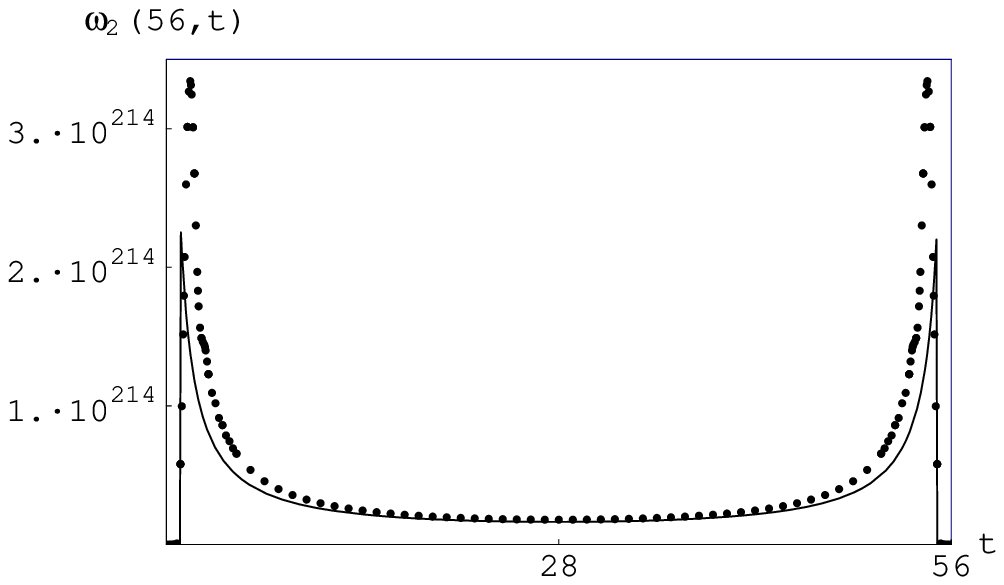}}}\fi
\caption{$\omega_2(56,t)$ computed numerically. The continuous line represents
$\omega_2$ as given by using $\Omega_1^h$ only ($\alpha' = 1$).}
\label{F1}}

\subsection{The calculation of Brandenberger and Vafa revisited}

Taking into account that\footnote{For $n=0$ we have that $\Omega_0\lp E \rp$
corresponds to the vacuum state, whose density of states is Dirac's delta.} 
$\Omega\lp E\rp=\sum_{n=0}^{\infty}\Omega_{n}\lp E\rp$, 
and using (\ref{single}), 
for the single string density of states when $E > \Lambda$,
Brandenberger and Vafa obtained

\be
\qquad \Omega^{\,h}\lp E\rp=
\frac{\mathrm{e}^{\bet E}}{2\pi E}\int_{-\infty}^{\infty}
\dif \alpha \,\mathrm{e}^{-\mathrm{i}\alpha}\,
\mathrm{e}^{\, \int_{\frac{\Lambda}{E}}^{1}
\frac{\dif x}{x}\mathrm{e}^{\mathrm{i}\alpha x}}
= \frac{\mathrm{e}^{\bet E}}{\Lambda}\lp a+b\,
\frac{\Lambda}{E}\rp
\label{omegavafa}
\ee
where:
\bes a&=&\frac{1}{2\pi}\int_{-\infty}^{\infty}\dif \alpha \,
\mathrm{e}^{\,\int_{0}^{\alpha}\dif x
\frac{\cos{x}-1}{x}}\, \cos{\lp  \int_{0}^{\alpha}\dif x
\frac{\sin{x}}{x}-\alpha\rp}=0.56\pm 0.01\\
b&=&\frac{1}{2\pi}\int_{-\infty}^{\infty}\alpha\,\dif \alpha \,
\mathrm{e}^{\,\int_{0}^{\alpha}\dif x
\frac{\cos{x}-1}{x}}\sin{\lp  \int_{0}^{\alpha}
\dif x\frac{\sin{x}}{x}-\alpha\rp}=-0.29\pm 0.01
\ees
The values of $a$ and $b$ were calculated numerically. Note that, 
although the exponent gives, when $\alpha\gg 1$, a factor $-\log\alpha$, 
the second integral is not convergent because of the oscillatory behavior of the integrand, 
and so it has to be regulated and they did it by using an exponentially decaying
factor. 
The sign of $b$ produces
a positive specific heat, and the Hagedorn temperature is then 
a maximum one in the microcanonical ensemble. 
We think, nevertheless, that this scenario needs a revision: there is 
nothing a priori unphysical in negative specific 
heats in the microcanonical ensemble, and  anyhow, 
we are not sure that there could be a positive specific heat phase in 
the microcanonical ensemble for the gas of strings with windings and momenta.

In fact, it is easy to see how a contradiction appears, when $b=-0.29$, 
in the following way: using \eqref{omegavafa} and the fact that 
$Z\lp\beta\rp={\cal{L}}\lk\Omega\lp E\rp\rk$ the following expression for
 the partition function can be obtained\footnote{The multiple string density 
of states must have a cutoff that indicates the range of validity of the asymptotic 
approximation and that would be, in general, different from the single string cutoff. 
We are using $\phi$ as the cutoff for the multiple string density of states.}:

\be 
Z^{\,h}\lp\beta\rp= \frac{a\,
\mathrm{e}^{-\lp\beta-\bet\rp \phi}}{\Lambda\lp\beta-\bet\rp}
+b\,\Gamma\lc 0,\phi\lp\beta-\bet\rp\rc\label{za}
\ee
whereas from the single density of states it is possible to arrive also to an 
expression for the multiple string partition function via the
equality 
$-\beta F\lp\beta\rp={\cal{L}}\lk\Omega_{1}\lp E\rp\rk$

\be 
Z^{\,h}\lp\beta\rp =\mathrm{e}^{-\beta F\lp\beta\rp}=\frac{
\mathrm{e}^{-\gamma}}{\Lambda\lp\beta-\bet\rp}+\mathrm{e}^{-\gamma}
+\mathrm{O}\lc\lp\beta-\bet\rp^1\rc\label{zgamma}
\ee
Comparing (\ref{za}) and (\ref{zgamma}), one immediately gets 
that\footnote{This expression for $a$ was roughly deduced in \cite{tan}, 
although no connection with the numerical value of \cite{vafa} was made there.}
 $a=\exp{\lp-\gamma\rp}=0.5614$, which perfectly agrees with the numerical 
value given in \cite{vafa}. But it is a very notorious fact that in
\eqref{zgamma} 
no term is found analogous to the logarithmically divergent one hidden in 
the incomplete gamma function of \eqref{za}.

 Furthermore, we have 
found an alternative calculation where the coefficient $b$ is actually zero and then, 
only one divergent term is present in both equations \eqref{za} and
\eqref{zgamma}. 
As it is written in \eqref{single}, the high energy dominant term in 
the single string density of states has a 
dimensionless factor which equals unity in the type II and the heterotic string 
(this factor could be volume dependent when considering open strings and branes). 
However, it is very useful to introduce in $\Omega_{1}\lp E\rp$ a factor $c$ 
that could be thought just as a regulator (whereas we are going to give it 
a physical meaning at the end of this  subsection). This change adds a factor
$c^{n}$ to 
$\Omega_{n}$ in \eqref{omegabase}. Only at the end the limit $c$ approaching to
one will be taken.
Furthermore, it is possible to work out the values of $a$ and $b$ analytically
if a simple 
change is made in \eqref{omegabase}: the key ingredient is noting that
 the upper limit in the integrals can be taken to infinity since the Dirac delta 
function ensures that no value greater than $E$ will contribute to them. 
This will render the final integrals much easier to perform; then, we will have

\be
\int_{\Lambda}^{E}\dif E_{i}\longrightarrow\int_{\Lambda}^{+\infty}\dif E_{i}
\qquad\Rightarrow\qquad
\Omega^{\,h}\lp E\rp=\frac{\mathrm{e}^{\bet E}}{2\pi
E}\int_{-\infty}^{+\infty}\dif
\alpha\,
 \mathrm{e}^{-\mathrm{i}\alpha}\,\mathrm{e}^{c \int_{\frac{\Lambda}{E}}^{\infty}
\frac{\dif x}{x}\mathrm{e}^{\mathrm{i}\alpha x}}
\ee

The calculation is now analogous to the one made in \cite{vafa},
 taking the first terms in the series expansion in $\Lambda/E$.

\ber
\Omega^{\,h}\lp E\rp&=& \frac{\mathrm{e}^{\bet E}}{2\pi
E}\int_{-\infty}^{+\infty}
\dif \alpha\,\mathrm{e}^{-\mathrm{i}\alpha}\,
\mathrm{e}^{c\int_{\frac{\alpha\Lambda}{E}}^{\infty\cdot\mathrm{sg}\lp\alpha\rp}
\frac{\dif x}{x}\cos{x}}\,\mathrm{e}^{\mathrm{i} 
c\int_{\frac{\alpha \Lambda}{E}}^{\infty\cdot\mathrm{sg}\lp\alpha\rp}
\frac{\dif x}{x}\sin{x}}\nonumber\\&
=&\frac{\mathrm{e}^{\bet E}}{2\pi E}\int_{-\infty}^{+\infty}\dif \alpha\,
\mathrm{e}^{-c\cdot\mathrm{Ci}\lp\frac{\alpha\Lambda}{E}\rp}\,
\cos\lp\alpha+c\cdot\mathrm{si}\lp\frac{\alpha\Lambda}{E}\rp\rp
\label{omegaintegral}
\eer
 Where $\mathrm{sg}\lp \alpha\rp$ stands for the sign function.
It is important to note that the $\mathrm{Ci}$ and the $\mathrm{si}$ functions
only coincide with the standard cosine integral and sine integral functions
for positive values of $\alpha$. $\mathrm{Ci}\lp x\rp$ is a real, even
function 
of $x$ whereas $\mathrm{si}\lp x\rp$ is a real, odd function of its argument 
with a discontinuity at $x=0$ and $\mathrm{si}\lp 0\rp= 0$. This is easy to
understand 
as a consequence of the integral upper limits including the $\mathrm{sg}\lp\alpha\rp$
 term in the exponential factors in \eqref{omegaintegral}. 
The integrand is then even and one could perform a series expansion in terms 
of $\alpha\Lambda/E$. 
Approximating $\mathrm{Ci}\lp\frac{\alpha\Lambda}{E}\rp=\gamma+\log\lp\frac{\alpha\Lambda}
{E}\rp+\mathrm{O}\lp\frac{\alpha\Lambda}{E}\rp$ and 
$\mathrm{si}\lp\frac{\alpha\Lambda}{E}\rp= \frac{-\pi}{2}+
\frac{\alpha\Lambda}{E}+\mathrm{O}\lp\frac{\alpha\Lambda}{E}\rp^{2}
\,(\alpha>0)$, we arrive at

\be
\Omega^{\,h}\lp E\rp=\mathrm{e}^{-\gamma}\,\frac{\mathrm{e}^{\bet E}}{\pi E}\,
\lp\frac{E}{\Lambda}\rp^{c}\int_{0}^{\infty}\frac{\dif\alpha}{\alpha^{c}}
\cos\lp\alpha-c\frac{\pi}{2}+c\frac{\alpha\Lambda}{E}\rp
\ee 

Using that $\cos\lp a-b\rp=\cos a \cos b+\sin a\sin b$ and
taking the lowest order in $\alpha\Lambda/E$ one finally gets

\be
\Omega^{\,h}\lp E\rp=
\frac{E^{c-1}}{\Lambda^{c}}\,\mathrm{e}^{\bet E}\lc a(c)+
 b(c)\,\frac{\Lambda}{E}+\mathrm{O}\lp \frac{\Lambda}{E}\rp^{2}\,\rc
\label{omegac}\ee
\bes 
a\lp c\rp&=&\frac{\mathrm{e}^{-c\gamma}}{\pi}\int_{0}^{\infty}\frac{\dif\alpha}
{\alpha^{c}}\cos\lp\frac{c\pi}{2}-\alpha\rp =
\frac{\mathrm{e}^{-c\gamma}}{\Gamma\lp c\rp}\qquad\mbox{if $0<c<1$}\\
b\lp c\rp&=&\frac{c\,\mathrm{e}^{-c\gamma}}{\pi}\int_{0}^{\infty}\frac{\dif\alpha}
{\alpha^{c-1}}\sin\lp\frac{c\pi}{2}-\alpha\rp=\frac{c\,\mathrm{e}^{-c\gamma}}
{\Gamma\lp c-1\rp}\qquad \mbox{if $1<c<2$}
\ees
The integrals converge only for the indicated values of\footnote{One could be 
tempted to put directly, in the expression for $a\lp c\rp$, that,
 when $c=1$, $\cos\lp\frac{\pi}{2}-\alpha\rp=\sin\alpha$; 
but this would produce a wrong result since we would be 
forgetting that $\mathrm{si}\lp 0\rp=0$.}  $c$.
Clearly, in the $c\rightarrow 1$ limit, the result for $a$ is fully 
compatible with the numerical value given in \cite{vafa}; the problem is
that it is very easy to see how in this limit $b$ goes to zero.  
As a matter of fact this method
can be generalized and more terms of the form $d_n \lp
\Lambda/E\rp^{n} \, (n\in \mathbb{N})$  can be computed, 
giving all of them zero for $d_n$ in the $c \rightarrow 1$ limit.

Now, it is straightforward to see how the contradiction between 
the equations  analogous to \eqref{za} and \eqref{zgamma}, that now depend 
on $c$, has disappeared. Using 
$\Omega_{1}\lp E\rp=c\,\mathrm{e}^{\beta_{\scriptscriptstyle H} E}/E$
we have that

\be
Z^{\,h}\lp\beta\rp=\mathrm{e}^{-c\,\beta	F^{\,h}\lp\beta\rp}\approx\frac{
\mathrm{e}^{-c\,\gamma}}{\Lambda^{c}\lp\beta-\bet\rp^{c}}+
\frac{c\,\mathrm{e}^{-c\,\gamma}}{\Lambda^{c-1}\lp\beta-\bet\rp^{c-1}}+
\mathrm{O} \lp\lp\beta-\bet\rp^{2-c}\rp.
\ee
And making directly the Laplace transform of \eqref{omegac}, 
the same expression for $Z\lp\beta\rp={\cal{L}}\lk \Omega\lp E\rp\rk$ 
can be found for $c>1$. When $c=1$ an annoying constant term 
appears preventing us to fix more than the only divergent term.

Once the value $c=1$ is taken, the expression for the 
density of states of the string gas in a finite size container is given by

\be
\Omega^{\,h}\lp E\rp= \mathrm{e}^{-\gamma}\,
\frac{\mathrm{e}^{\bet E}}{\Lambda}
\qquad E\gg\Lambda.
\ee

From this we can calculate both the entropy of the system and its
temperature.  The fundamental thermodynamic relationship giving the
entropy as a function of the energy now looks like

\be
S^{h}\lp E\rp=\bet E+\log\lp\frac{\mathrm{e}^{-\gamma}}{\Lambda}\rp.
\label{S}
\ee
Comparing it with \eqref{entropic}, the analogous expression in the fixed
temperature,
case we see how both ensembles seem to be  inequivalent as pointed in
\eqref{fluc}.
With this entropy we will also have that temperature is fixed to
Hagedorn's, and we would have to conclude that $C_V\lp E \rp $ would be
infinite.

The constant $c$ has been introduced as a mere way of doing
analytical continuation of ill-defined expressions, but we can give it a
physical interpretation. Lets look at the expression defining $\Omega\lp E
\rp$ once $c$ is introduced

\be
\Omega\lp E\rp =\sum_{n=0}^{\infty}c^{n}\,\Omega_{n}\lp E\rp
\ee

We see that $c$ is really acting as the fugacity of the system, so that
working with a generic value of $c$ and then performing the $c \rightarrow
1$ limit is exactly the same as working with a generic non null chemical
potential and then taking  it to zero.  This interpretation also lets us know
that we have not been working in the microcanonical ensemble, but in the
"enthalpic" one \cite{exten} for which energy and the chemical potential are
given. This way $\Omega\lp E\rp$ now depends on $c$ 
and becomes $\Omega\lp E, c\rp$.

\subsection{Fluctuations for the fixed energy description}

It is now clear that the density of states of the string gas is given by

\be
\Omega^{\,h}\lp E\rp = \frac{\mathrm{e}^{-\gamma}}{\Lambda}\,\theta\lp E
-\phi\rp\,\mathrm{e}^{\bh \,E}
\label{density}
\ee
The multi-string energy cutoff $\phi$ cannot be
completely determined without matching the high energy regime with the low 
energy phase because, imposing that $\beta F^{\,h}\lp \beta\rp$ must be
obtained, $\phi$ would appear as a factor of $\lp\beta -\beta_H\rp^{\,0}$ that
is a regular term being $F^{l}\lp\beta\rp$  also regular at $\beta_H$. The
matching can be done, but we are not going to dwell further on this point.

An immediate consequence of \eqref{S}
is that the high energy microcanonical specific
heat, $C_V^{\,h } \lp E\rp$, is divergent contrary to what, by using
\cite{vafa}, has been assumed as
true for more than fifteen years\footnote{In the next subsection and the
conclusions, we will treat and discuss
the relevance that the radius corrections can have in relation to this
thermodynamical statement.}. 

Once $\Omega\lp E,c\rp$ for the enthalpic ensemble is obtained, it is
straightforward to compute 
the number of strings and its fluctuations as

\be
\begin{split}
\overline{N}\lp E \rp &= \left. c~\partial_{c}\log \Omega\lp
E,c\rp \right|_{c=1}\approx\log\lp\frac{E}{\Lambda}\rp
\nonumber\\
\frac{\sqrt{\overline{\Delta N^{2}}\lp E \rp }}{\overline{N} \lp E \rp }&=
\frac{\sqrt{\left. \lp c~\partial_{c} \rp ^2 \log
\Omega\lp E,c\rp  
\right|_{c=1}}}{\overline{N} \lp E \rp }
\approx \frac{1}{\sqrt{\overline{N}\lp E \rp}}
\end{split}
\ee
with
 
\be
\begin{split}
\Omega^{\,h}\lp E,c\rp = \theta
\lp E-\phi \rp\frac{\mathrm{e}^{\,-c\gamma}}{\Lambda^{\,c}\Gamma\lp c\rp} &
E^{\,c-1} \mathrm{e}^{\bh E}\,\\ & \lc 1+
  \lp c-1\rp\,\mathrm{O}\lp\frac{\Lambda}{E}\rp + ...\rc
\end{split}
\ee 
that has been obtained by Laplace inversion of
$\mathrm{e}^{\,c\,q\lp\beta\rp}$.

\subsection{Other refinements}

In the preceding sections any dependence on $R$ has been lost as a result of
being in a physical situation in which sums can be well approximated 
by integrals. Now,  we
would like to add the effects of
introducing Euler-Maclaurin corrections  \cite{nuestro} in the integrals that
represent sums
over
windings and momenta. This can also be done by means of  a Poisson resummation.
As a result, there appear more singular points
 in $-\beta F\lp\beta \rp$ whose location depends on  $R$ \cite{tan}.

\begin{equation}
-\beta F^{\,h}\lp\beta\rp =
\sum_{\vec{m},\vec{n},j} \,\Gamma\lc 0, \frac{2\pi\,\lp\beta^{\,2}
-\beta_{\vec{m},\vec{n},j}\lp R \rp ^{\,2}\rp}{\epsilon\beta_H^{\,2}}\rc
\end{equation}
with 
\be
\beta^{\,2}_{\vec{m},\vec{n},j} = \frac{\alpha' \pi^{\,2}}{\lp j+1/2
\rp^{\,2}}\,
\lc 2-\frac{R^ {\,2}}{\alpha'} \vec{m}^{\,2}- \frac{\alpha'}{R^{\,2}}
\vec{n}^{\,2} \rc
\ee
Assuming $ R >\sqrt{\alpha'}$ the
singularity nearest to $\bh$ is  $\beta_1=\bh-\eta$, $\eta
\approx \alpha'\bh/(4 R^{\,2}) $. 
Considering also the term depending on $\beta_1$  we get a
corrected expression for the density of states

\be
\Omega^{\,h}\lp E, R\rp = \frac{\mathrm{e}^{-\gamma}}{\Lambda}\,\theta\lp E
-\phi\rp\,\mathrm{e}^{\bh \,E}
 \lp 1- \frac{\mathrm{e}^{-\eta
E} \lp \eta E \rp ^{17} }{\Gamma \lp 18 \rp} \rp 
\lp\frac{\mathrm{e}^{-\gamma}}{2 \eta \Lambda} \rp ^{\,18}
\label{radius}
\ee
Where  $\eta E \gg 1$ has been  assumed.  This
expression depends on $R$ through $\eta$ and the $R$ corrections are actually
non extensive. If taken into account,  the effect of
these corrections on  $\beta$ would be
\be
\beta\lp E \rp =\beta_H +\eta \frac{\mathrm{e}^{
-\eta
E}}{\Gamma\lp 18 \rp} \, \lp \eta E \rp ^{17} 
\label{betaradius}
\ee 

The second term on the right hand side of this expression has already been
studied in the literature \cite{tan}, \cite{barbon} and has been frequently
claimed as being the cause of having a positive specific heat. 
It is easy to see that, after all, having enough energy for a given radius,  the
gas can always be in the regime in
which there is no dependence on the volume and this is the regime of the
Brandenberger-Vafa scenario, for which the specific heat diverges. The radius
corrections are exponentially suppressed and, for our system, are not different
from the finite volume corrections that are dropped when the thermodynamic limit
is taken over the ideal gas of particles (see, for example, \cite{exten}).

\section{Conclusions}

 As a first important result, it is crucial to remark that no
published calculation has found the $ -0.29\,\, \mathrm{e}^{\beta_H\,E}/E$ term
but the one
presented by  Brandenberger and Vafa in \cite{vafa}. This is the term that is
needed to state, as these authors do, that the microcanonical specific heat
is positive in the physical situation in which energy density is so high that
the density of states does not depend on the volume. As far as we know, 
what any
other calculation actually gets is that, for the same physical situation,
the
specific heat  is divergent because a  null $b$ coefficient is found (see
eq. \eqref{omegavafa}). In reference \cite{tan}, it is explicitly admitted that
$b=-0.29$ is not found, but the authors do not face the important question
that the contradiction between their calculation and that of Brandenberger and
Vafa rises. It seems that this is so because they consider they are using what
they call a different "formalism". In our work, we clearly show that, using the
same technique
 Brandenberger and Vafa used and a physically meaningful
regularization\footnote{Brandenberger and Vafa stay that they use an exponential
regulator for a numerical computation.}, the $b$
coefficient vanishes (and also any other high energy correction).
This is the first result of our work and, in our opinion, critically depends on
understanding that the "micro" ensemble is really the
"enthalpic" one, namely, a fixed energy and fixed chemical potential
ensemble.

We have used the $S$-representation of the Helmholtz free energy to finally
conclude that the behavior of the free energy around $\beta_H$ coincides with 
what is gotten from the $F$ representation. This might be expected
 but it is not a trivial fact\footnote{In the previous versions of our work we
thought on
the contrary because we found only the leading order contribution in
\eqref{tau1}. In fact, a seemed very concerned referee, after reading the first
version of our work, was fully convinced of the difference between the $F$ and
 $S$ representations for this calculation and used
it to reject the work without any hesitation because he/she liked more the
$F$-representation.},
 because both representations do not coincide. That the $S$ and $F$
representations are
not equal seems to be clearly commented for the first time in
\cite{kutasovseiberg}. The relationship between both representations is
carefully studied for the family of the heterotic strings in
\cite{mavazmaro} (see also \cite{mavazmaroplb1}). 
In those works it is clearly established  that
the 
$F$-representation does not provide an analytical function for complex
$\beta$. In other words,
 it is false that the $F$-representation can be seen as providing the analytical
continuation of the $S$-representation for heterotic strings. Then, it is not
rigorous to say that the $F$-representation of the free energy can be used to
get the density of states by inverse Laplace transformation of the
corresponding partition function. What one can only do is to continue the
$S$-representation. When one uses the $F$-representation to get the behavior
around $\beta_H$ one is really using it in the interval $(\beta_H, +\infty)$
where it coincides with the $S$-representation. 
Now, we are also providing a very concrete and explicit example of to
what extent the $S$ and
$F$ representations of the free energy are equal.

From other point of view, taking into account that 
the $F$-representation gives the free energy as computed for the 
compactification of the Euclidean time on a circle of length $\beta$ (including
string windings along it), in \cite{mpl} there appears a proof for
the noncritical $c=1$ string of
the recently emphasized fact that, for strings, the free energy
cannot be computed, at any temperature, by the compactification of
time \cite{seiberg}.

Another important point is to what extent the $R$ corrections presented 
in subsection (4.3) can be taken as showing that the system really has
a positive specific heat. This seems to be  the belief as expressed in
\cite{tan} and \cite{barbon}. In our opinion, the problem is that these
nonextensive corrections
are exponentially suppressed with the value of the  energy and are then
of no thermodynamical relevance for the system in the thermodynamical limit
in which the  energy density is very high and the sum over windings and momenta
can be replaced by an integral. Those corrections are as the finite 
volume $1/\sqrt{V}$ corrections for the gas of free particles; they are
irrelevant in the thermodynamical limit in which the sum over momenta can be
replaced by an integral.

In the case we treat the gas in the fixed temperature (canonical) description,
the big  fluctuations might justify the exclusion of the nonextensive term that
renders the canonical specific heat positive. In any case,  those
fluctuations would just
make the canonical equilibrium description physically unusable.
 
From a cosmological point of view, one could think that there would not be any
relevant cosmological implication
from our results because, after all, the equation of state would still be the
same, corresponding to pressureless dust matter ($P=0$). However, things are
more intricate because a divergent microcanonical specific heat can be an
indication of a
phase transition. In our case, what we have done in the microcanonical
(really enthalpic) treatment is a description of how the volume of phase
space  in the N-body problem changes when energy, which is a conserved quantity,
increases \cite{gross}. We have found that the system behaves very differently
from the
grand canonical ensemble in which $T_H$ would be a maximum temperature and the
specific heat, as a function of temperature, would be positive. 

What is clear is that a divergent microcanonical specific heat cannot be used
ab initio as a criterion to drop as unphysical our gas of closed strings at
finite size.

\section*{Acknowledgments}
The work of M. A. Cobas is partially supported by a Spanish  MEC-FPI fellowship.
M. Su\'arez is partially supported by a  Spanish MEC-FPU fellowship. We all are
partially supported by the Spanish MEC project BFM2003-00313.

\section*{A. The UV limit in the S-representation partition function}
\setcounter{equation}{0}
\renewcommand{\theequation}{A.\arabic{equation}}

This appendix is devoted to explain how \eqref{tau1} is obtained and to
explicitly show that
\eqref{magic} holds.

First of all, the integral computing the sum over windings and momenta can be
performed to give

\be
\int_{\mathbb{R}^9\times\mathbb{R}^9}\, \dif\vec{l}\,\dif\vec{k}\, 
 \mathrm{e}^{-\pi\tau_2(\vec{l}^{\,2} + \vec{k}^{\,2})}\mathrm{e}^
{2\pi \mathrm{i}\,\vec{l}\cdot\vec{k}\tau_1} = \left |\tau\right|^{-9}
\ee

Next, it has to be noticed that, when $\tau\rightarrow 0$, it holds that
 $$\left|\theta_2 (0,-1/(2\tau))\right|^{\,-16}
\approx 2^{\,-16} \,\mathrm{e}^{2\pi\tau_2/\left|\tau\right|^{\,2}}$$ because
all the other terms are finite when $\tau\rightarrow 0$.
One has then to perform the integral over $\tau_1$ as
providing a function of $\tau_2$ given by  $\mathrm{e}^{\,2\pi/\tau_2}$ times a
series expansion in powers of $\tau_2$ as it appears on the right hand side
of \eqref{tau1}.
Namely, the left hand side of \eqref{tau1} (we called it $I(\tau_2)$) is now
given by

\be
\begin{split}
I(\tau_2) =
\frac{1}{2^{\,8} \tau_2}\,\mathrm{e}^{2\pi/\tau_2}\,
\int_{-1/2}^{+1/2}\,\,
\dif\tau_1\,\,\lp 1+ \frac{\tau_1^{\,2}}{\tau_2^{\,2}}\rp^{-1/2}
\mathrm{e}^{-2\pi\tau_1^{\,2}/\tau_2^{\,3}}
\,\,\mathrm{e}^{\frac{2\pi}{\tau_2\lc 1+ \tau_1^{\,2}/\tau_2^{\,2}\rc} -
\frac{2\pi}{\tau_2} + \frac{2\pi\tau_1^{\,2}}{\tau_2^{\,3}}}
\end{split}
\ee
It has been written in a way prepared to be rewritten in terms of a function
$\tilde{I}(\tau_2)$ as $I(\tau_2) =
2^{\,-8}\tau_2^{\,1/2}\mathrm{e}^{\,2\pi/\tau_2}\,\tilde{I}(\tau_2)$ where

\be
\tilde{I}(\tau_2) = 
\int_{-\lp 2\tau_2^{3/2}\rp^{-1}}^{\lp 2\tau_2^{3/2}\rp^{-1}}\, \dif x\,
\,\,\mathrm{e}^{-2\pi x^{\,2}}\,
\mathrm{e}^{\frac{2\pi}{\tau_2}\lc
\frac{1}{1+x^{\,2}\tau_2} - \lp 1-\tau_2 x^{\,2}\rp\rc}\lp 1+ \tau_2
x^{\,2}\rp^{-1/2}
\ee
For which the change of variables $\tau_1= x\,\tau_2^{3/2}$ has been used.
Now, for the product of the last two factors in the integrand,
the following series expansion can be written

\be
\begin{split}
\mathrm{e}^{\frac{2\pi}{\tau_2}\lc
\frac{1}{1+x^{\,2}\tau_2} - \lp 1-\tau_2 x^{\,2}\rp\rc} & \lp 1+\tau_2
x^{\,2}\rp^{-1/2} = \mathrm{e}^{2\pi x^{\,4}\tau_2 / (1+\tau_2
x^{\,2})}\lp 1+\tau_2 x^{\,2}\rp^{-1/2}\\\
& = \sum_{b=0}^{+\infty}\,
\sum_{a=0}^{+\infty}\,\,
\frac{(-1)^{\,a}(2\pi)^{\,b}\,\Gamma\lp b+a +1/2\rp\,}{b!\,\,a!\,\, \Gamma\lp b
+ 1/2\rp}\,\tau_2^{\,b + a}\, x^{\,4b + 2a}
\end{split}
\ee

Next we are able  to perform the integral over $x$ of the term $x^{4b+2a}$
obtaining

\be
\begin{split}
\int_{-\lp 2\tau_2^{3/2}\rp^{-1}}^{\lp 2\tau_2^{3/2}\rp^{-1}}\, \dif x\,
\mathrm{e}^{-2\pi x^{\,2}}\, x^{4b + 2a} & =
\frac{\Gamma\lp 2b+a + 1/2\rp - \Gamma\lp 2b+a+1/2,
2\pi/(4\tau_2^{\,3})\rp}{(2\pi)^{2b+a+1/2}} \\ &\approx 
(2\pi)^{-2b-a-1/2}\,\Gamma\lp
2b+ a + 1/2\rp
\end{split}
 \ee
 where the last approximation results from the exponential suppression of the
contribution coming from the incomplete Gamma function when its argument gets
big (what
happens here when $\tau_2\rightarrow 0$).

Now one of the sums in the resulting double sum to get $\tilde{I}$ can be
calculated (!!) to finally give

\be
\tilde{I}(\tau_2) = \sum_{b=0}^{+\infty}\, (2\pi)^{-b-1/2}\,\Gamma\lp
b+1/2\rp\,\tau_2^{\,b}
\ee 

We are then able to get $I(\tau_2)$ and, in particular, the coefficients
$a_i$ as

\be
a_i = \frac{\Gamma\lp i+1/2\rp}{(2\pi)^{\,i}\sqrt{\pi}}
\ee
The next step is to use $I(\tau_2)$ as written in \eqref{tau1} with the
already known $a_i$ coefficients to compute $F^{\,h}(\beta)$ finding then
the $b_n$ factors in \eqref{libre}. This is easily done to give

\be
b_n = (-1)^{\,n}\,\frac{a_n}{\beta_H\,n!}
\ee
 
The final computation is that of the series generated by the $b_n$ coefficients
as it appears on the left hand side of \eqref{magic}, namely

\be
R(\beta)\equiv\sum_{n=0}^{+\infty}\, b_n \lp \beta^{\,2}
-\beta_H^{\,2}\rp^{n}\lp \frac{2\pi}{\beta_H^{\,2}}\rp^{n}
\ee
Taking into account that $\beta^2 -\beta_H^{\,2} = 2\beta_H\lp \beta
-\beta_H\rp\lp 1 + \frac{\beta -\beta_H}{2\beta_H}\rp$, $R(\beta)$ can be
written as a double sum

\be
R(\beta) = \sum_{n=0}^{+\infty}\,\lc
\frac{1}{\sqrt{\pi}}\,\sum_{q=0}^{n}\,\frac{(-1)^{\,q} \,2^{\,2q-n}\Gamma\lp
q+1/2\rp}{\lp n-q\rp! \lp 2q-n\rp!}\rc \,\,\frac{\lp \beta -
\beta_H\rp^{n}}{\beta_H^{\,n+1}}
\ee
The term between square brackets can be computed to give exactly $(-1)^{\,n}$.
So, we have finally showed that 

\be
R(\beta) = 1/\beta
\ee
by showing that it is given by the power series expansion of $1/\beta$ around
$\beta_H$.

\end{document}